\documentclass[12pt]{article}
\usepackage{graphicx}
\usepackage{amsmath}
\usepackage{amssymb}
\usepackage{caption2}
\begin{document}
\bibliographystyle{prsty}
\begin{center}
{\large {\bf \sc{   Strong decay $\Delta^{++} \to
p\pi$ with light-cone QCD sum rules  }}} \\[2mm]
Zhi-Gang Wang \footnote{E-mail:wangzgyiti@yahoo.com.cn.  }   \\
  Department of Physics, North China Electric Power University, Baoding 071003, P. R. China
 \end{center}

\begin{abstract}
In this article, we calculate the strong coupling constant
$g_{\Delta N\pi}$ and study  the strong decay $\Delta^{++} \to p\pi$
with the light-cone QCD sum rules. The numerical value of the strong
coupling constant $g_{\Delta N\pi}$ is consistent with  the
experimental data.  The small discrepancy maybe due to failure to
take into account the perturbative $\mathcal {O}(\alpha_s)$
corrections.
\end{abstract}


\section{Introduction}

The $\Delta(1232)$ resonance dominates many nuclear phenomena at
energies above the pion-production threshold and plays an  important
role in the physics of the strong interaction. It is almost an ideal
elastic $\pi N$ resonance, and decays into the nucleon  and pion
($\Delta \to N\pi$) with the branching  fraction  about $99\%$. The
only other (electromagnetic) decay channel ($\Delta \to N\gamma$)
contributes less than $1\%$ to the total decay width \cite{PDG}.
There is a very small mass gap (less than $300\rm{MeV}$) between the
$\Delta$ and the nucleon, and the $\Delta(1232)$ is taken as an
explicit dynamical degree of freedom in the heavy baryon chiral
perturbation theory \cite{HBCHPT}.

 In this article, we calculate the strong coupling constant $g_{\Delta N\pi}$
  with the light-cone QCD sum rules, and study the decay width $\Gamma_{\Delta \to N\pi}$. The strong
 coupling constants of the octet baryons with the vector and pseudoscalar mesons
 $g_{ NNV}$ and $g_{NNP}$ have been calculated with the
 light-cone QCD sum rules \cite{Aliev06}.
 The light-cone QCD sum
rules  carry out the operator product expansion near the light-cone
$x^2\approx 0$ instead of the short distance $x\approx 0$ while the
nonperturbative hadronic matrix elements  are parameterized by the
light-cone distribution amplitudes   instead of  the vacuum
condensates \cite{LCSR,LCSRreview}. The nonperturbative
 parameters in the light-cone distribution amplitudes are calculated with  the conventional QCD  sum rules
 and the  values are universal \cite{SVZ79}.

The article is arranged as: in Section 2, we derive the strong
coupling constant  $g_{\Delta N\pi}$ with the light-cone QCD sum
rules; in Section 3, the numerical result and discussion; and
Section 4 is reserved for conclusion.

\section{Strong coupling constant  $g_{\Delta N\pi}$ with light-cone QCD sum rules}

In the following, we write down the
 two-point correlation function $\Pi_\mu(p,q)$,
\begin{eqnarray}
\Pi_\mu(p,q)&=&i \int d^4x \, e^{-i q \cdot x} \,
\langle 0 |T\left\{ J_p(0)\bar{J}_\mu(x)\right\}|\pi(p)\rangle \, , \\
J_p(x)&=& \epsilon^{abc} u^T_a(x)C\gamma_\mu u_b(x) \gamma_5 \gamma^\mu d_c(x)\, ,  \nonumber \\
J_\mu(x)&=& \epsilon^{abc} u^T_a(x)C\gamma_\mu u_b(x)  u_c(x)\, ,
\end{eqnarray}
where the baryon currents $J_p(x)$ and $J_\mu(x)$
 interpolate the octet baryon $p$ and  decuplet baryon $\Delta^{++}$
 respectively \cite{Ioffe81}, the external  state $\pi$ has the
four momentum $p_\mu$ with $p^2=m_\pi^2$ . The general form of the proton
current can be written as \cite{Baryoncurrent}
\begin{eqnarray}
J_p(x,t) &=&  \epsilon_{abc} \left\{ \left[ u_a^T(x) C d_b(x)
\right] \gamma_5 u_c(x) + t  \left[ u_a^T(x) C \gamma_5 d_b(x)
\right]  u_c(x) \right\} \, ,\nonumber
\end{eqnarray}
in the limit $t=-1$, we recover the Ioffe current. If we retain  the
additional  parameter
  $t$ and choose the ideal value, the sum rule maybe improved, in this article, we choose the
Ioffe current for simplicity. The correlation function
$\Pi_{\mu}(p,q)$  can be decomposed as
\begin{eqnarray}
\Pi_{\mu }(p,q)&=&\Pi\sigma_{\alpha\beta}p^\alpha q^\beta
p_\mu+\Pi_{A1}p_\mu+\Pi_{A2}\not\!\!q p_\mu+\Pi_{A3}\not\!\!pp_\mu+
\nonumber\\
&&\Pi_{B1}q_\mu+\Pi_{B2}\not\!\!q
q_\mu+\Pi_{B3}\not\!\!pq_\mu+\Pi_{B4}\sigma_{\alpha\beta}p^\alpha
q^\beta
q_\mu+\nonumber\\
&&\Pi_{C1}\gamma_\mu+\Pi_{C2}\not\!\!q\gamma_\mu+\Pi_{C3}\not\!\!p\gamma_\mu+\Pi_{C4}\epsilon_{\mu\nu\alpha\beta}\gamma^\nu
\gamma_5 p^\alpha q^\beta
\end{eqnarray}
due to the Lorentz invariance, where the $\Pi$ and $\Pi_i$ are
Lorentz invariant functions of  $p$ and $q$. In this article, we
choose the tensor structure $\sigma_{\alpha\beta}p^\alpha q^\beta
p_\mu$ for analysis.

The strong coupling among the $\Delta$, $p$ and $\pi$ can be
described by the following chiral Lagrangian \cite{HBCHPT},
\begin{eqnarray}
 \mathcal {L}(x)&=&g_{\Delta N\pi} \left[\bar{\Delta}^\mu(x)\partial_\mu \pi(x)N(x)+
 \bar{N}(x)\partial_\mu \pi(x)\Delta^\mu(x)\right] \, .
 \end{eqnarray}

Basing on the quark-hadron duality \cite{SVZ79}, we can insert  a
complete series of intermediate states with the same quantum numbers
as the current operators $J_p(x)$    and $J_\mu(x)$ into the
correlation function $\Pi_{\mu}(p,q)$  to obtain the hadronic
representation. After isolating the ground state contributions from
the pole terms of the baryons $p$    and $\Delta$, we get the
following result\footnote{In the first version of this
article(arXiv:0707.3736), the numerical factor $\frac{1}{3}$ is
missed in $\frac{g_{\Delta N\pi}}{3}\sigma_{\alpha\beta}p_\alpha
q_\beta p_\mu $ and we obtain too small value for the strong
coupling constant $g_{\Delta N \pi}$ to accommodate   the
experimental data.} ,
\begin{eqnarray}
\Pi_{\mu }(p,q)&=&\frac{\langle0| J_p(0)| N(q+p)\rangle\langle
N(q+p)| \Delta(q) \pi(p) \rangle \langle \Delta(q)|\bar{J}_\mu(0)|
0\rangle}
  {\left\{M_{p}^2-(q+p)^2\right\}\left\{M_{\Delta}^2-q^2\right\}}  + \cdots \nonumber \\
&=& \frac{\lambda_p
\lambda_\Delta}{\left\{M_p^2-(q+p)^2\right\}\left\{M_\Delta^2-q^2\right\}}
\left\{ \frac{g_{\Delta N\pi}}{3}\sigma_{\alpha\beta}p_\alpha
q_\beta p_\mu+\cdots\right\}+\cdots \, ,
\end{eqnarray}
where the following definitions have been used,
\begin{eqnarray}
\langle 0| J_p (0)|N(p)\rangle &=&\lambda_p U(p,s) \, , \nonumber \\
\langle 0| J_\mu (0)|\Delta(p)\rangle &=&\lambda_\Delta U_\mu(p,s) \, , \nonumber \\
\sum_sU(p,s)\overline {U}(p,s)&=&\!\not\!{p}+M_p \, , \nonumber \\
\sum_s U_\mu(p,s) \overline{U}_\nu(p,s)
&=&-(\!\not\!{p}+M_{\Delta})\left\{ g_{\mu\nu}-\frac{\gamma_\mu
\gamma_\nu}{3}-\frac{2p_\mu p_\nu}{3M_{\Delta}^2}+\frac{p_\mu
\gamma_\nu-p_\nu \gamma_\mu}{3M_{\Delta}} \right\} \, , \nonumber\\
\langle N(q')| \Delta(q) \pi(p) \rangle&=&ig_{\Delta N
\pi}\overline{U}(q',s')U_\mu(q,s)p^\mu\, ,
\end{eqnarray}
the last identity corresponds to the phenomenological Lagrangian in
Eq.(4).

 The current $J_\mu(x)$ couples
not only to the isospin $I=\frac{3}{2}$ and spin-parity $J^P=\frac{3}{2}^+$ states,
but also to the isospin $I=\frac{3}{2}$ and spin-parity $J^P=\frac{1}{2}^-$ states.
 For a generic $\frac{1}{2}^-$ resonance   $\Delta^*$ \cite{Braun06},
\begin{eqnarray}
 \langle0|J_{\mu}(0)|\Delta^*(p)\rangle=\lambda_{*}
 (\gamma_{\mu}-4\frac{p_{\mu}}{M_{*}})U^{*}(p,s) \, ,
\end{eqnarray}
where $\lambda^{*}$ is  the  pole residue  and $M_{*}$ is the mass.
The spinor $U^*(p,s)$  satisfies the usual Dirac equation
$(\not\!\!p-M_{*})U^{*}(p)=0$. If we take the phenomenological
Lagrangian,
\begin{eqnarray}
\mathcal {L}(x)&=& g_{\Delta^* N \pi}\left\{\bar{\Delta}^*(x) N(x)
\pi(x) +\bar{N}(x) \Delta^*(x) \pi(x) \right\} \, ,
\end{eqnarray}
which corresponds to $\langle N(q')| \Delta^*(q) \pi(p)
\rangle=g_{\Delta^* N \pi}\overline{U}(q',s')U^*(q,s)$,  the
contributions from the $\frac{1}{2}^-$ states can be written as
\begin{eqnarray}
\Pi_\mu(p,q)&=&\frac{g_{\Delta^* N \pi}\lambda_p
\lambda_*}{\left\{M_p^2-(q+p)^2\right\}\left\{M_*^2-q^2\right\}}
\left\{ (\not\!\!p+\not\!\!q +M_p)(\not\!\!q +M_*)(\gamma_{\mu}-4\frac{q_{\mu}}{M_{*}})\right\}\nonumber\\
&&+\cdots \nonumber\\
&=&\Pi_{D}\not\!\!q p_\mu +\Pi_{E1}q_\mu+\Pi_{E2}\not\!\!q
q_\mu+\Pi_{E3}\not\!\!pq_\mu+\Pi_{E4}\sigma_{\alpha\beta}p^\alpha
q^\beta
q_\mu+\nonumber\\
&&\Pi_{F1}\gamma_\mu+\Pi_{F2}\not\!\!q\gamma_\mu+\Pi_{F3}\not\!\!p\gamma_\mu+\Pi_{F4}\epsilon_{\mu\nu\alpha\beta}\gamma^\nu
\gamma_5 p^\alpha q^\beta \, ,
\end{eqnarray}
where the  $\Pi_i$ are Lorentz invariant functions of  $p$ and $q$.
 If we choose the tensor structure $\sigma_{\alpha\beta}p_\alpha q_\beta
p_\mu$, the $\Delta^*$  has no contaminations.

In the following, we briefly outline the  operator product expansion
for the correlation function $\Pi_{\mu }(p,q)$  in perturbative QCD
theory. The calculations are performed at the large space-like
momentum regions $(q+p)^2\ll 0$  and $q^2\ll 0$, which correspond to
the small light-cone distance $x^2\approx 0$ required by the
validity of the operator product expansion approach. We write down
the "full" propagator of a massive light  quark in the presence of
the quark and gluon condensates firstly \cite{LCSR,SVZ79}\footnote{One
can consult the first article of Ref.\cite{LCSR}  and  the second article of Ref.\cite{SVZ79} for the technical
details in deriving the full propagator.},
\begin{eqnarray}
S_{ab}(x) &=&
\frac{i\delta_{ab}\!\not\!{x}}{ 2\pi^2x^4}
-\frac{\delta_{ab}m_q}{4\pi^2x^2}-\frac{\delta_{ab}}{12}\langle
\bar{q}q\rangle +\frac{i\delta_{ab}}{48}m_q
\langle\bar{q}q\rangle-\frac{\delta_{ab}x^2}{192}\langle
\bar{q}g_s\sigma Gq\rangle
\nonumber\\
&& +\frac{i\delta_{ab}x^2}{1152}m_q\langle \bar{q}g_s\sigma
Gq\rangle \!\not\!{x}\nonumber\\
&&-\frac{i}{16\pi^2x^2} \int_0^1 dv
\left[(1-v)g_sG_{\mu\nu}(vx)\!\not\!{x}
\sigma^{\mu\nu}+vg_sG_{\mu\nu}(vx)\sigma^{\mu\nu}
\!\not\!{x}\right]  \nonumber\\
&&+\cdots \, ,
\end{eqnarray}
then contract the quark fields in the correlation function
$\Pi_\mu(p,q)$ with the Wick theorem, and obtain the following result,
\begin{eqnarray}
\Pi_\mu(p,q)&=&2i\epsilon_{abc}\epsilon_{a'b'c'} \int d^4x  e^{-i q
\cdot x} \nonumber \\
&&\left\{ Tr\left[ \gamma_\alpha  S_{bb'}(-x)\gamma_\mu CS^T_{aa'}(-x)C\right]\gamma_5\gamma^\alpha \langle 0|d_c(0)\bar{u}_{c'}(x) |\pi(p)\rangle\right.\nonumber \\
&&\left. -2\gamma_5 \gamma^\alpha \langle
0|d_c(0)\bar{u}_{b'}(x)|\pi(p)\rangle\gamma_\mu C S^T_{aa'}(-x)C
\gamma_\alpha S_{bc'}(-x)\right\}\, .
\end{eqnarray}
Perform the following Fierz re-ordering to extract the contributions
from the two-particle and three-particle $\pi$-meson light-cone
distribution amplitudes respectively,
\begin{eqnarray}
q^a_\alpha(0) \bar{q}^b_\beta(x)&=&-\frac{1}{12}
\delta_{ab}\delta_{\alpha\beta}\bar{q}(x)q(0)
-\frac{1}{12}\delta_{ab}(\gamma^\mu)_{\alpha\beta}\bar{q}(x)\gamma_\mu
q(0) \nonumber\\
&&-\frac{1}{24}\delta_{ab}(\sigma^{\mu\nu})_{\alpha\beta}\bar{q}(x)\sigma_{\mu\nu}q(0) \nonumber\\
&&+\frac{1}{12}\delta_{ab}(\gamma^\mu
\gamma_5)_{\alpha\beta}\bar{q}(x)\gamma_\mu \gamma_5 q(0)\nonumber\\
&&+\frac{1}{12}\delta_{ab}(i \gamma_5)_{\alpha\beta}\bar{q}(x)i
\gamma_5 q(0) \, , \\
q^a_\alpha(0) \bar{q}^b_\beta(x)G^{ba}_{\lambda\tau}(vx)&=&-\frac{1}{4}
 \delta_{\alpha\beta}\bar{q}(x)G_{\lambda\tau}(vx)q(0)
-\frac{1}{4} (\gamma^\mu)_{\alpha\beta}\bar{q}(x)\gamma_\mu G_{\lambda\tau}(vx)
q(0) \nonumber\\
&&-\frac{1}{8} (\sigma^{\mu\nu})_{\alpha\beta}\bar{q}(x)\sigma_{\mu\nu}G_{\lambda\tau}(vx)q(0) \nonumber\\
&&+\frac{1}{4} (\gamma^\mu
\gamma_5)_{\alpha\beta}\bar{q}(x)\gamma_\mu \gamma_5 G_{\lambda\tau}(vx)q(0)\nonumber\\
&&+\frac{1}{4} (i \gamma_5)_{\alpha\beta}\bar{q}(x)i \gamma_5G_{\lambda\tau}(vx) q(0)
\, ,
\end{eqnarray}
 and substitute the hadronic matrix elements (such as the $ \langle
0| {\bar u} (x) \gamma_\mu \gamma_5 d(0) |\pi(p)\rangle$,  $ \langle
0| {\bar u} (x)g_s\sigma_{\mu\nu}\gamma_5 G_{\alpha\beta}(vx) d(0)
|\pi(p)\rangle$, $ \langle 0| {\bar u} (x) \sigma_{\mu\nu}\gamma_5
d(0) |\pi(p)\rangle$, etc.)  with
 the corresponding $\pi$-meson light-cone distribution amplitudes\footnote{ In calculations, we have used the relations
 $\sigma_{\mu\nu}=-\frac{i}{2}\epsilon_{\mu\nu\alpha\beta}\sigma^{\alpha\beta}\gamma_5$ and
  $\widetilde{G}_{\mu\nu}=\frac{1}{2}\epsilon_{\mu\nu\alpha\beta}G^{\alpha\beta}$.}, finally we
obtain the spectral density  at the coordinate space. Once the spectral
density in the coordinate space is obtained,
 we can translate it  into the
 momentum space with the $D=4+2\epsilon$ dimensional Fourier transform,

\begin{eqnarray}
4\Pi&=&\frac{2f_\pi}{3\pi^2}
 \int_0^1 du u\phi_{\pi}(u)\frac{\Gamma(\epsilon)}{(-Q^2)^\epsilon}-\frac{f_\pi m_\pi^2}{2\pi^2}
 \int_0^1 du u A(u)\frac{\Gamma(1)}{(-Q^2)^{1}}
\nonumber \\
&&+\frac{f_\pi}{9}\langle\frac{
\alpha_sGG}{\pi}\rangle
 \int_0^1 du u\phi_{\pi}(u)\frac{\Gamma(2)}{(-Q^2)^{2}}\nonumber \\
&&-\frac{f_\pi m_\pi^2}{36}\langle\frac{
\alpha_sGG}{\pi}\rangle\int_0^1 du u A(u)\frac{\Gamma(3)}{(-Q^2)^{3}}
\nonumber \\
&& +\frac{f_{3\pi}\langle \bar{q}q\rangle}{2}\int_0^1 dvv \int_0^1 d\alpha_g \int_0^{1-\alpha_g} d\alpha_u\nonumber \\
&&   \frac{\Gamma(2)}{(-Q^2)^{2}}\mid_{u=\alpha_u+v\alpha_g} \phi_{3\pi}(\alpha_u,\alpha_g,1-\alpha_u-\alpha_g)\nonumber \\
&&+\frac{f_\pi m_\pi^2}{2\pi^2} \int_0^1 dv \int_0^1 d\alpha_g \int_0^{1-\alpha_g} d\alpha_u u \frac{\Gamma(1)}{(-Q^2)^{1}}\mid_{u=\alpha_u+v\alpha_g}\nonumber \\
&& \left[(1-9v)V_{\perp}-4(1-2v)A_{\parallel}-4(1-v)A_{\perp} \right](\alpha_u,\alpha_g,1-\alpha_u-\alpha_g) \nonumber \\
&&-\frac{4f_\pi m_\pi^2}{\pi^2}
\int_0^1 dv \int^1_0 d\alpha_g \int_0^{1-\alpha_g} d\alpha_u
\int_0^{\alpha_u} d\alpha  \frac{\Gamma( 1)}{(-Q^2)^{ 1}}\mid_{u=\alpha_u+v\alpha_g} \nonumber\\
&&\left[V_{\parallel}+V_{\perp}+(1-2v)(A_{\parallel}+A_{\perp})
\right](\alpha,\alpha_g,1-\alpha-\alpha_g) \nonumber\\
&&+\frac{4f_\pi m_\pi^2}{\pi^2}
\int_0^1 dv (1-v)\int^1_0 d\alpha_g \int_0^{\alpha_g} d\beta
\int_0^{\beta} d\alpha  \frac{\Gamma( 1)}{(-Q^2)^{ 1}}\mid_{u=1-(1-v)\alpha_g}\nonumber\\
&&\left[V_{\parallel}+V_{\perp}+(1-2v)(A_{\parallel}+A_{\perp})
\right](\alpha,\beta,1-\alpha-\beta) \, ,
\end{eqnarray}
where $Q_\mu=q_\mu+up_\mu$ and
$Q^2=(1-u)q^2+u(p+q)^2-u(1-u)m_\pi^2$. The $\epsilon$ is a small
positive quantity, after taking the double Borel transform, we can
take the limit $\epsilon\rightarrow 0$.

There is no contribution from terms  of the form $\langle
\bar{q}q\rangle \phi_\pi(u)$, while there is rather large
contribution from that terms in the sum rules for the strong
coupling constant $g_{NN\pi}$, see the article "V. M. Braun and I.
E. Filyanov, Z. Phys.  {\bf C44} (1989) 157" in Ref.\cite{LCSR}.  If
we replace the decuplet baryon current $J_\mu(x)$ with the octet
baryon current $J_n(x)$ (interpolating the neutron) and study the
strong coupling constant $g_{NN\pi}$,
 the Feynman diagrams
 are quite different. Our mathematica code can be used to calculate
the strong coupling constant $g_{NN\pi}$ and produce the terms
$\langle \bar{q}q\rangle \phi_\pi(u)$.

The decuplet baryon current $J_\mu(x)$ and octet  baryon current
$J_p(x)$ have the Dirac structures $\gamma_\mu \otimes 1$ and
$\gamma_\alpha \otimes \gamma^\alpha \gamma_5$ respectively, where
$\otimes$ stands for the $u$ quark fields. The Dirac structure
$\gamma_\alpha \gamma_5$ corresponds to the twist-2 light-cone
distribution amplitude $\phi_\pi(u)$. If we replace one of the
"full" $u$ quark propagators with the quark condensate $\langle
\bar{q}q\rangle$, the terms $\langle \bar{q}q\rangle \phi_\pi(u)$ in
the correlation function $\Pi_{\mu}(p,q)$ have the Dirac structures
$\gamma_\mu \gamma_\alpha \gamma_\lambda $ or $\gamma_\mu
\gamma_\lambda\gamma_\alpha
 $, which are chiral even, because only the
perturbative part of the other "full" $u$ quark propagator has
contribution. It is not unexpected that they have no contribution to
the chiral odd structure $\sigma_{\alpha\beta}p_\alpha q_\beta
p_\mu$.

 The light-cone
distribution amplitudes $\phi_{\pi}(u)$, $A(u)$,
$\phi_{3\pi}(\alpha_i)$, $A_\perp(\alpha_i)$,
$A_\parallel(\alpha_i)$, $V_\perp(\alpha_i)$ and
$V_\parallel(\alpha_i)$ of the $\pi$ meson are presented in the
appendix \cite{PSLC}, the nonperturbative parameters in the
light-cone distribution amplitudes are scale dependent, in this
article, the energy scale is taken to be $\mu=1\,\rm{GeV}$. The
contributions proportional to the $G_{\mu\nu}$ can give rise to
three-particle (and four-particle) meson distribution amplitudes
with a gluon (and quark-antiquark pair) in addition to the two
valence quarks, their corrections are usually not expected to play
any significant roles\footnote{For examples, in the decay $B \to
\chi_{c0}K$, the factorizable contribution is zero and the
nonfactorizable contributions from the soft hadronic matrix elements
are too small to accommodate the experimental data \cite{WangLH};
the net contributions from the three-valence particle light-cone
distribution amplitudes to the strong coupling constant
$g_{D_{s1}D^*K}$ are rather small, about $20\%$ \cite{Wang0611}. In
Ref.\cite{Wang07}, we observe that the contributions from the
three-particle (quark-antiquark-gluon) light-cone distribution
amplitudes are less than $5\%$ for the strong coupling constants
$g_{D^*D^*P}$. In this article, the contributions from the
three-particle light-cone distribution amplitudes are about $10\%$.
The contributions from the three-particle (quark-antiquark-gluon)
distribution amplitudes of the mesons are always of minor importance
comparing with the two-particle (quark-antiquark) distribution
amplitudes in the light-cone QCD sum rules.   In our previous work,
we also study the four form-factors $f_1(Q^2)$, $f_2(Q^2)$,
$g_1(Q^2)$ and $g_2(Q^2)$ of the $\Sigma \to n$ with the light-cone
QCD sum rules up to twist-6 three-quark light-cone distribution
amplitudes and obtain satisfactory results \cite{Wang06}.  In a
word, we can neglect the contributions from the valence gluons and
make relatively rough estimations
 in the light-cone QCD sum rules. }. In this article, we take them into account for completeness.

Taking double Borel transform  with respect to the variables
$Q_1^2=-q^2$ and $Q_2^2=-(p+q)^2$ respectively (i.e.
$\frac{\Gamma[n]}{\left[u(1-u)m_\pi^2+(1-u)Q_1^2+uQ_2^2\right]^n}
 \rightarrow \frac{M^{2(2-n)}}{M_1^2 M_2^2} e^{-\frac{u(1-u)m_\pi^2}{M^2}}
\delta(u-u_0)$, $M^2= \frac{M^2_1M^2_2}{M^2_1+M^2_2}$ and $u_0=
\frac{M_1^2}{M_1^2+M_2^2}$),  then subtract the contributions from
the high resonances and continuum states by introducing  the
threshold parameter $s_0$ (i.e. $ M^{2n}\rightarrow
\frac{1}{\Gamma[n]}\int_0^{s_0} ds s^{n-1}e^{-\frac{s}{M^2}}$),
finally we obtain the  sum rule for the strong coupling constant
$g_{\Delta N \pi}$,
\begin{eqnarray}
g_{\Delta N \pi}&=&\frac{3}{\lambda_p \lambda_\Delta}
\exp\left\{\frac{M_\Delta^2}{M_1^2}+\frac{M_p^2}{M_2^2}-\frac{u_0(1-u_0)m_\pi^2}{M^2}\right\}
\left\{\frac{u_0}{6\pi^2} M^4E_1(x)
 f_\pi\phi_{\pi}(u_0) \right.
\nonumber \\
&&-\frac{u_0}{8\pi^2}M^2E_0(x)f_\pi
 m_\pi^2 A(u_0)+\frac{u_0}{36 }\langle\frac{ \alpha_sGG}{\pi}\rangle f_\pi
  \phi_{\pi}(u_0)\nonumber \\
&&-\frac{u_0}{144 }\langle\frac{ \alpha_sGG}{\pi}\rangle\frac{f_\pi
 m_\pi^2A(u_0)}{M^2}
\nonumber \\
&& +\frac{1}{8}\langle \bar{q}q\rangle
f_{3\pi}   \int_0^{u_0} d\alpha_u \int_{u_0-\alpha_u}^{1-\alpha_u} d\alpha_g \frac{u_0-\alpha_u}{\alpha_g^2}
\phi_{3\pi}(\alpha_u,\alpha_g,1-\alpha_u-\alpha_g)\nonumber \\
&&-\frac{u_0}{8\pi^2}M^2E_0(x)f_\pi m_\pi^2 \int_0^{u_0} d\alpha_u
\int_{u_0-\alpha_u}^{1-\alpha_u} d\alpha_g \frac{1}{\alpha_g}
 \nonumber\\
&&\left[4(1-2\frac{u_0-\alpha_u}{\alpha_g})A_{\parallel}+4(1-\frac{u_0-\alpha_u}{\alpha_g})A_{\perp}\right.\nonumber\\
&&\left.-(1-9\frac{u_0-\alpha_u}{\alpha_g})V_{\perp}\right](\alpha_u,\alpha_g,1-\alpha_u-\alpha_g) \nonumber \\
&&-\frac{1}{\pi^2}  M^2E_0(x)f_\pi m_\pi^2 \left[\int_0^{1-u_0}
d\alpha_g \int^{u_0}_{u_0-\alpha_g} d\alpha_u \int_0^{\alpha_u}
d\alpha \right.\nonumber\\
&&\left.+\int^1_{1-u_0} d\alpha_g \int^{1-\alpha_g}_{u_0-\alpha_g}
d\alpha_u
\int_0^{\alpha_u} d\alpha\right]\frac{1}{\alpha_g} \nonumber\\
&&\left[V_{\parallel}+V_{\perp}+(1-2\frac{u_0-\alpha_u}{\alpha_g})(A_{\parallel}+A_{\perp})
\right](\alpha,\alpha_g,1-\alpha-\alpha_g) \nonumber\\
&&+\frac{1}{\pi^2}  M^2E_0(x)f_\pi m_\pi^2 (1-u_0)\int_{1-u_0}^1
d\alpha_g\frac{1}{\alpha_g^2} \int_0^{\alpha_g}d\beta
\int_0^{1-\beta}d\alpha\nonumber\\
&&\left.\left[V_{\parallel}+V_{\perp}+(1-2\frac{u_0+\alpha_g-1}{\alpha_g})(A_{\parallel}+A_{\perp})
\right](\alpha,\beta,1-\alpha-\beta)\right\} \, ,
\end{eqnarray}
where
\begin{eqnarray}
E_n(x)&=&1-(1+x+\frac{x^2}{2!}+\cdots+\frac{x^n}{n!})e^{-x} \, , \nonumber\\
x&=&\frac{s_0}{M^2} \, .\nonumber
\end{eqnarray}

In the following, we present an ansatz for the spectral density at
the level of quark-gluon degrees of freedom \cite{Kim03,LiZH02}.
Firstly, we perform a double Borel transform for the correlation
function (which is  denoted as $\int_0^1
du\frac{\Gamma(\alpha)f(u)}{[u(1-u)m_\pi^2+(1-u)Q_1^2+uQ_2^2]^\alpha}$
symbolically) with respect to the variables $Q_{1}^{2}$ and
$Q_{2}^{2}$ respectively, and obtain the result,
\begin{eqnarray}
&&B_{M_2}B_{M_1}\int_0^1
du\frac{\Gamma(\alpha)f(u)}{[u(1-u)m_\pi^2+(1-u)Q_1^2+uQ_2^2]^\alpha} \nonumber\\
&=&\frac{M^{2(2-\alpha)}}{M_{1}^{2}M_{2}^{2}}\exp\left[-\frac{u_0(1-u_0)m_\pi^2}
{M^{2}} \right]f( u_{0})  \, ,
\end{eqnarray}
where the $f(u)$ stand for the light-cone distribution amplitudes,
$\alpha <2$, $u_0=\frac{M_1^2}{M_1^2+M_2^2}$,
$M^2=\frac{M_1^2M_2^2}{M_1^2+M_2^2}$. Then we introduce the
corresponding spectral densities $\rho(s_1,s_2)$,
\begin{eqnarray}
&&M^{2(2-\alpha)} \exp\left[-\frac{u_0(1-u_0)m_\pi^2} {M^{2}}
\right]f( u_{0})\nonumber\\
 &=& \int_0^{\infty}ds_1
\int_0^{\infty}ds_2 \exp\left[-\frac{s_1} {M_1^{2}}-\frac{s_2}
{M_2^{2}} \right] \rho(s_1,s_2) \, ,
\end{eqnarray}
 and take a replacement $M_{1}^{2}\rightarrow \frac{1}{\sigma
_1}$, $M_{2}^{2}\rightarrow \frac{1}{\sigma _2}$,
\begin{eqnarray}
&&\int_0^{\infty}ds_1 \int_0^{\infty}ds_2
 \exp\left\{-s_1\sigma_1-s_2\sigma_2 \right\}
\rho(s_1,s_2) \nonumber\\
 &=&\frac{f( u_{0})}{(\sigma_1
+\sigma_2)^{2-\alpha}}\exp\left\{-u_0(1-u_0)m_\pi^2(\sigma_1+\sigma_2)\right\}\nonumber\\
&=&\frac{f( u_{0})}{\Gamma(2-\alpha)}\int_0^\infty d\lambda
\lambda^{1-\alpha}
\exp\left\{-\left[u_0(1-u_0)m_\pi^2+\lambda\right](\sigma_1+\sigma_2)\right\}
\,.
\end{eqnarray}
Finally we take a double Borel transform with respect to the
variables  $ \sigma _{1}$ and $ \sigma _{2}$ respectively, the
resulting QCD spectral densities read
\begin{eqnarray}
\rho(s_1,s_2)&=&\frac{f(u_0)}{\Gamma(2-\alpha)}\left\{s_1-u_0(1-u_0)m_\pi^2\right\}^{1-\alpha}\delta(s_1-s_2)
\, .
\end{eqnarray}

The threshold parameter $s_0$ is taken as
$s_0=\rm{max}(s_1^0,s^0_2)$, where the $s_1^0$ and $s_2^0$ are the
threshold parameters for the channels $1$ and $2$ respectively.
 The  quantity $u_0(1-u_0)m_\pi^2$ is tiny and can
be safely neglected.  Our approach (i.e. performing a double Borel
transform and taking a replacement $ M^{2n}\rightarrow
\frac{1}{\Gamma[n]}\int_0^{s_0} ds s^{n-1}e^{-\frac{s}{M^2}}$.) is
an indirect way to obtain the same results.

\section{Numerical result and discussion}

The input parameters are taken as
$m_u=m_d=(0.0056\pm0.0016)\,\rm{GeV}$, $f_\pi=0.130\,\rm{GeV}$,
$m_{\pi} =0.138\,\rm{GeV}$, $\lambda_3=0.0$,
$f_{3\pi}=(0.45\pm0.15)\times 10^{-2}\,\rm{GeV}^2$,
$\omega_3=-1.5\pm0.7$, $\omega_4=0.2\pm0.1$, $a_2=0.25\pm 0.15$,
$a_1=0.0 $, $\eta_4=10.0\pm3.0$ \cite{LCSR,PSLC,Belyaev94}, $\langle
\bar{q}q \rangle=-(0.24\pm 0.01\, \rm{GeV})^3$, $\langle
\frac{\alpha GG}{\pi}\rangle=(0.33\,\rm{GeV})^4 $ \cite{SVZ79},
$M_{p}=0.938\,\rm{GeV}$, $M_{\Delta}=1.232\,\rm{GeV}$,
$\lambda_{p}=(2.4\pm 0.2)\times 10^{-2}\,\rm{GeV}^3$ and
$\lambda_{\Delta}=(3.0\pm0.2)\times 10^{-2}\,\rm{GeV}^3$
\cite{Ioffe81}.

In this article, we neglect  the  perturbative $\mathcal
{O}(\alpha_s)$ corrections to the strong coupling constant
$g_{\Delta N \pi}$, and take the values of the pole residues
$\lambda_{p}$ and $\lambda_{\Delta}$ without perturbative $\mathcal
{O}(\alpha_s)$ corrections  for consistency.

In calculation, we observe  the main uncertainties come from the two
parameters $a_2$ and $\eta_4$ in the two-particle light-cone
distribution amplitudes, as the dominant  contributions come from
the two-particle light-cone distribution amplitudes $\phi_\pi(u)$
and $A(u)$, the contributions from the terms involving the
three-particle (quark-antiquark-gluon) light-cone distribution
amplitudes  are of minor importance, about $7\%$ of the contribution
from the term $\frac{2u_0}{3\pi^2}M^4E_1(x)f_\pi
 \phi_{\pi}(u_0)$. The
uncertainty of the parameter $a_2$ obtained in Ref.\cite{PSLC} is
very large, in this article, we take smaller uncertainty, say $30\%$
(i.e. $a_2=0.25\pm 0.08$),  which is the typical uncertainty in the
QCD sum rules.

The values of the vacuum condensates have been updated with the
experimental data for the $\tau$ decays, the QCD sum rules for the
baryon masses and analysis of the charmonium spectrum
\cite{Zyablyuk, Ioffe2005,AlphaS},  in this article, we choose the
standard (or old) values to keep in consistent with the sum rules
used in determining the non-perturbative parameters in the
light-cone distribution amplitudes.

The threshold parameter $s_0$ is chosen to be
$s_0=(3.4\pm0.1)\,\rm{GeV}^2$ to avoid possible contamination from
the contribution of the $P_{33}$ baryon $\Delta(1920)$ in the
$p\pi^+$ scattering amplitude \cite{Arndt2006}. Furthermore, it is
large enough to take into account  the contribution of the
$\Delta(1232)$. However, the interpolating current $J_\mu(x)$
 has nonvanishing
coupling with the isospin $I=\frac{3}{2}$ and spin $J=\frac{1}{2}$ states, the contribution from the
 $S_{31}$ state $\Delta(1620)$ is included in if the $\Delta(1620)$ has negative parity
 \cite{PDG}. We choose the tensor structure
$\sigma_{\alpha\beta}p_\alpha q_\beta p_\mu$ to avoid the
contamination.

 The Borel parameters are chosen as $\frac{M_\Delta^2}{M_1^2}=\frac{M_p^2}{M_2^2}$ and
 $M^2=\frac{M_1^2M^2_p}{M_\Delta^2+M^2_p}=(2.0-3.0)\,\rm{GeV}^2$,
 in those regions,
the value of the strong coupling constant $g_{\Delta N\pi}$ is
rather stable with the variation of the Borel parameter $M^2$, which
are shown in  Fig.1.

Taking into account all the uncertainties, finally we obtain the
numerical results for the strong coupling constant $g_{\Delta N
\pi}$, which are shown in  Fig.1,
\begin{eqnarray}
  g_{\Delta N \pi} &=&(13.5\pm 7.2)\,\rm{GeV}^{-1} \, \, , \nonumber \\
  g_{\Delta N \pi} &=&(13.5\pm 5.4)\,\rm{GeV}^{-1} \, \, ,
\end{eqnarray}
for the parameter  $a_2=0.25\pm0.15$ and     $a_2=0.25\pm0.08$
respectively.

\begin{figure}
\centering
  \includegraphics[totalheight=6cm,width=7cm]{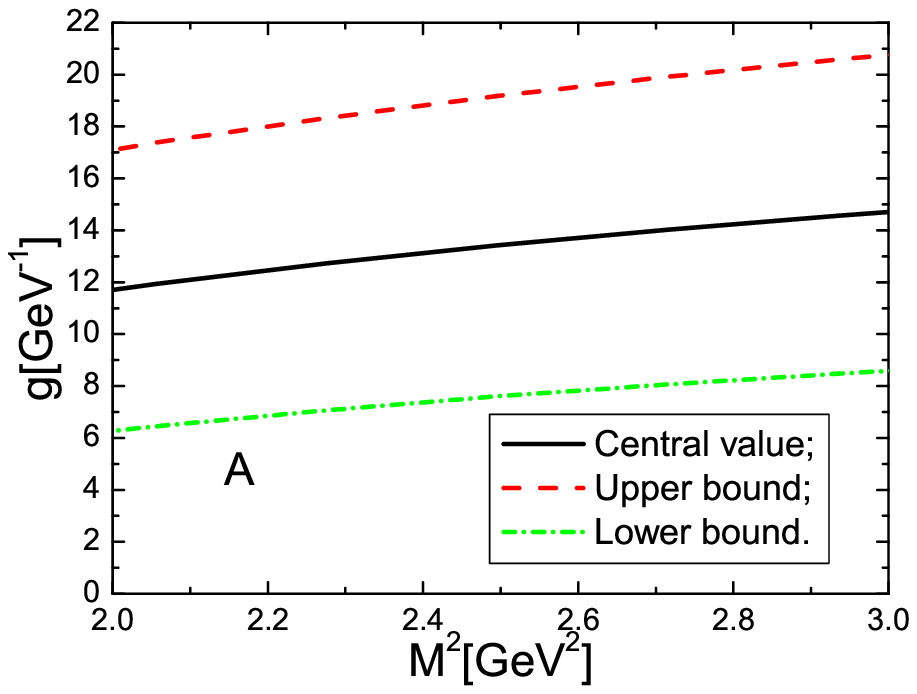}
   \includegraphics[totalheight=6cm,width=7cm]{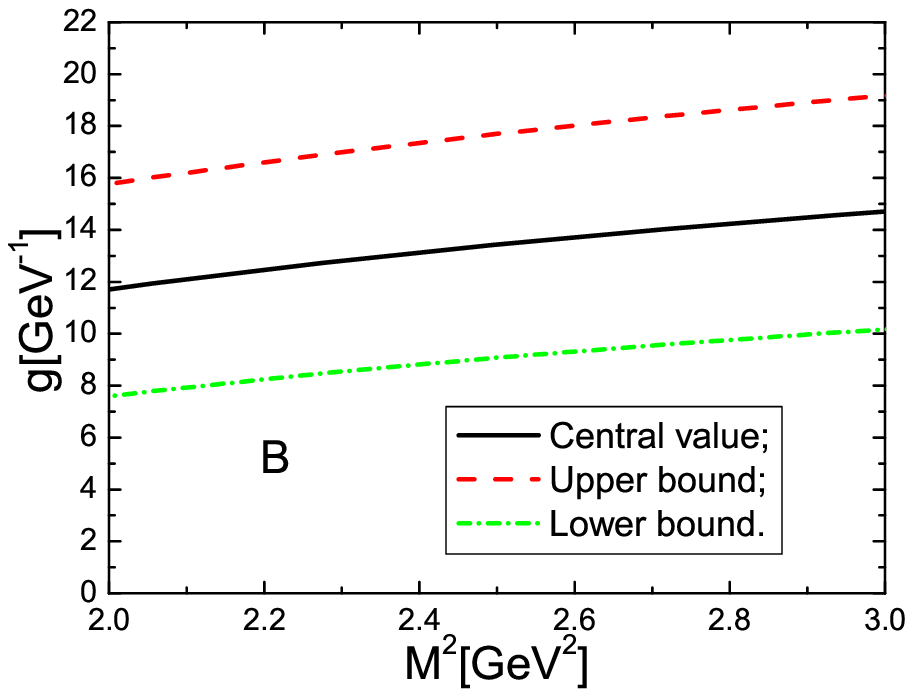}
           \caption{ $g_{\Delta N\pi}$ with the Borel parameter $M^2$, $A$ for $a_2=0.25\pm0.15$ and $B$ for $a_2=0.25\pm0.08$. }
\end{figure}

The strong coupling constant $g_{\Delta N\pi}$ has the following
relation with the decay width $\Gamma_{\Delta \to N\pi}$,
\begin{eqnarray}
\Gamma_{\Delta \to N\pi}&=&\frac{g^2_{\Delta N\pi} p_{cm}}{32\pi
M_\Delta^2} \sum_{ss'} \mid \overline{U}(p',s) p^\mu_\pi
U_\mu(p'',s') \mid^2 \, ,
\nonumber \\
p_{cm}&=&\frac{\sqrt{[M_\Delta^2-(M_p+m_\pi)^2][M_\Delta^2-(M_p-m_\pi)^2]}}{2M_\Delta}
\, .
\end{eqnarray}
 If we take the experimental data as input
parameter, $\Gamma_{\Delta \to N\pi}=118\,\rm{GeV}$ \cite{PDG}, we
can obtain the value $g_{\Delta N \pi}\approx 15.6 \,\rm{GeV}^{-1}$,
our numerical result $g_{\Delta N \pi} =(13.5\pm
5.4)\,\rm{GeV}^{-1}$ is rather good.

In the region $M^2=(2.0-3.0)\,\rm{GeV}^2$,
$\frac{\alpha_s(M)}{\pi}\sim 0.10-0.13$ \cite{AlphaS}. If the
radiative $\mathcal {O}(\alpha_s)$ corrections to the leading
perturbative terms are companied  with large numerical factors, just
like in the case of the QCD sum rules for the mass of the proton
\cite{Ioffe2005},
\begin{eqnarray}
1+(\frac{53}{12}+\gamma_E) \frac{\alpha_s(M)}{\pi} \sim
1+(0.54-0.65) \, ,
\end{eqnarray}
the contributions of the order $\mathcal {O}(\alpha_s)$ are large,
neglecting them can impair the predictive ability.  Furthermore, the
pole residues $\lambda_p$ and $\lambda_\Delta$ also receive
contributions from the perturbative $\mathcal {O}(\alpha_s)$
corrections, if we take them into account properly, we can improve
the value of the strong coupling constant $g_{\Delta N \pi}$.

\section{Conclusion}

In this article, we calculate the strong coupling constant
$g_{\Delta N\pi}$  and study the strong  decay $\Delta^{++} \to p
\pi^+$ with the light-cone QCD sum rules. The numerical value of the
strong coupling constant $g_{\Delta N\pi}$ is consistent with  the
experimental data.  The small discrepancy maybe due to failure to
take into account the perturbative $\mathcal {O}(\alpha_s)$
corrections.
 \section*{Appendix}
 The light-cone distribution amplitudes of the $\pi$ meson are defined
 by \cite{PSLC}
\begin{eqnarray}
\langle0| {\bar u} (x) \gamma_\mu \gamma_5 d(0) |\pi(p)\rangle& =& i
f_\pi p_\mu \int_0^1 du  e^{-i u p\cdot x}
\left\{\phi_\pi(u)+\frac{m_\pi^2x^2}{16}
A(u)\right\}\nonumber\\
&&+\frac{i}{2}f_\pi m_\pi^2\frac{x_\mu}{p\cdot x}
\int_0^1 du  e^{-i u p \cdot x} B(u) \, , \nonumber\\
\langle0| {\bar u} (x) i \gamma_5 d(0) |\pi(p)\rangle &=&
\frac{f_\pi m_\pi^2}{
m_u+m_d}\int_0^1 du  e^{-i u p \cdot x} \phi_p(u)  \, ,  \nonumber\\
\langle0| {\bar u} (x) \sigma_{\mu \nu} \gamma_5 d(0) |\pi(p)\rangle
&=&i(x_\mu p_\nu-x_\nu p_\mu)  \frac{f_\pi m_\pi^2}{6 (m_u+m_d)}
\int_0^1 du
e^{-i u p \cdot x} \phi_\sigma(u) \, ,  \nonumber\\
\langle0| {\bar u} (x) \sigma_{\mu \nu} \gamma_5 g_s G_{\alpha \beta
}(v x)d(0) |\pi(p)\rangle&=& f_{3 \pi}\left\{(p_\mu p_\alpha
g^\bot_{\nu
\beta}-p_\nu p_\alpha g^\bot_{\mu \beta}) -(p_\mu p_\beta g^\bot_{\nu \alpha}\right.\nonumber\\
&&\left.-p_\nu p_\beta g^\bot_{\mu \alpha})\right\} \int {\cal
D}\alpha_i \phi_{3\pi} (\alpha_i)
e^{-ip \cdot x(\alpha_u+v \alpha_g)} \, ,\nonumber\\
\langle0| {\bar u} (x) \gamma_{\mu} \gamma_5 g_s G_{\alpha
\beta}(vx)d(0) |\pi(p)\rangle&=&  f_\pi m_\pi^2p_\mu  \frac{p_\alpha
x_\beta-p_\beta x_\alpha}{p
\cdot x}\nonumber\\
&&\int{\cal D}\alpha_i A_{\parallel}(\alpha_i) e^{-ip\cdot
x(\alpha_u +v \alpha_g)}\nonumber \\
&&+ f_\pi m_\pi^2 (p_\beta g^\perp_{\alpha\mu}-p_\alpha
g^\perp_{\beta\mu})\nonumber\\
&&\int{\cal D}\alpha_i A_{\perp}(\alpha_i)
e^{-ip\cdot x(\alpha_u +v \alpha_g)} \, ,  \nonumber\\
\langle0| {\bar u} (x) \gamma_{\mu} i g_s \tilde G_{\alpha
\beta}(vx)d(0) |\pi(p)\rangle&=& f_\pi m_\pi^2 p_\mu  \frac{p_\alpha
x_\beta-p_\beta x_\alpha}{p \cdot
x}\nonumber\\
&&\int{\cal D}\alpha_i V_{\parallel}(\alpha_i) e^{-ip\cdot
x(\alpha_u +v \alpha_g)}\nonumber \\
&&+ f_\pi m_\pi^2 (p_\beta g^\perp_{\alpha\mu}-p_\alpha g^\perp_{\beta\mu})\nonumber\\
&&\int{\cal D}\alpha_i V_{\perp}(\alpha_i) e^{-ip\cdot x(\alpha_u +v
\alpha_g)} \, ,
\end{eqnarray}
where $g_{\mu\nu}^\perp=g_{\mu\nu}-\frac{p_\mu x_\nu+p_\nu x_\mu}{p
\cdot x}$, $\tilde G_{\mu \nu}= \frac{1}{2} \epsilon_{\mu\nu
\alpha\beta} G^{\alpha\beta} $ and ${\cal{D}} \alpha_i =d \alpha_u d
\alpha_d d \alpha_g \delta(1-\alpha_u -\alpha_d -\alpha_g)$.

The light-cone distribution amplitudes of the $\pi$ meson are
parameterized as \cite{PSLC}
\begin{eqnarray}
\phi_\pi(u)&=&6u(1-u)
\left\{1+a_1C^{\frac{3}{2}}_1(\xi)+a_2C^{\frac{3}{2}}_2(\xi)
\right\}\, , \nonumber\\
\phi_p(u)&=&1+\left\{30\eta_3-\frac{5}{2}\rho^2\right\}C_2^{\frac{1}{2}}(\xi)\nonumber \\
&&+\left\{-3\eta_3\omega_3-\frac{27}{20}\rho^2-\frac{81}{10}\rho^2 a_2\right\}C_4^{\frac{1}{2}}(\xi)\, ,  \nonumber \\
\phi_\sigma(u)&=&6u(1-u)\left\{1
+\left[5\eta_3-\frac{1}{2}\eta_3\omega_3-\frac{7}{20}\rho^2-\frac{3}{5}\rho^2 a_2\right]C_2^{\frac{3}{2}}(\xi)\right\}\, , \nonumber \\
\phi_{3\pi}(\alpha_i) &=& 360 \alpha_u \alpha_d \alpha_g^2 \left \{1
+\lambda_3(\alpha_u-\alpha_d)+ \omega_3 \frac{1}{2} ( 7 \alpha_g
- 3) \right\} \, , \nonumber\\
V_{\parallel}(\alpha_i) &=& 120\alpha_u \alpha_d \alpha_g \left(
v_{00}+v_{10}(3\alpha_g-1)\right)\, ,
\nonumber \\
A_{\parallel}(\alpha_i) &=& 120 \alpha_u \alpha_d \alpha_g a_{10}
(\alpha_d-\alpha_u)\, ,
\nonumber\\
V_{\perp}(\alpha_i) &=& -30\alpha_g^2
\left\{h_{00}(1-\alpha_g)+h_{01}\left[\alpha_g(1-\alpha_g)-6\alpha_u
\alpha_d\right] \right.  \nonumber\\
&&\left. +h_{10}\left[
\alpha_g(1-\alpha_g)-\frac{3}{2}\left(\alpha_u^2+\alpha_d^2\right)\right]\right\}\,
, \nonumber\\
A_{\perp}(\alpha_i) &=&  30 \alpha_g^2 (\alpha_u-\alpha_d) \left\{h_{00}+h_{01}\alpha_g+\frac{1}{2}h_{10}(5\alpha_g-3)  \right\}, \nonumber\\
A(u)&=&6u(1-u)\left\{
\frac{16}{15}+\frac{24}{35}a_2+20\eta_3+\frac{20}{9}\eta_4 \right.
\nonumber \\
&&+\left[
-\frac{1}{15}+\frac{1}{16}-\frac{7}{27}\eta_3\omega_3-\frac{10}{27}\eta_4\right]C^{\frac{3}{2}}_2(\xi)
\nonumber\\
&&\left.+\left[
-\frac{11}{210}a_2-\frac{4}{135}\eta_3\omega_3\right]C^{\frac{3}{2}}_4(\xi)\right\}+\left\{
 -\frac{18}{5}a_2+21\eta_4\omega_4\right\} \nonumber\\
 && \left\{2u^3(10-15u+6u^2) \log u+2\bar{u}^3(10-15\bar{u}+6\bar{u}^2) \log \bar{u}
 \right. \nonumber\\
 &&\left. +u\bar{u}(2+13u\bar{u})\right\} \, ,\nonumber\\
 g(u)&=&1+g_2C^{\frac{1}{2}}_2(\xi)+g_4C^{\frac{1}{2}}_4(\xi)\, ,\nonumber\\
 B(u)&=&g(u)-\phi_\pi(u)\, ,
\end{eqnarray}
where
\begin{eqnarray}
h_{00}&=&v_{00}=-\frac{\eta_4}{3} \, ,\nonumber\\
a_{10}&=&\frac{21}{8}\eta_4 \omega_4-\frac{9}{20}a_2 \, ,\nonumber\\
v_{10}&=&\frac{21}{8}\eta_4 \omega_4 \, ,\nonumber\\
h_{01}&=&\frac{7}{4}\eta_4\omega_4-\frac{3}{20}a_2 \, ,\nonumber\\
h_{10}&=&\frac{7}{2}\eta_4\omega_4+\frac{3}{20}a_2 \, ,\nonumber\\
g_2&=&1+\frac{18}{7}a_2+60\eta_3+\frac{20}{3}\eta_4 \, ,\nonumber\\
g_4&=&-\frac{9}{28}a_2-6\eta_3\omega_3 \, ,
\end{eqnarray}
   $\xi=2u-1$, and $ C_2^{\frac{1}{2}}(\xi)$, $ C_4^{\frac{1}{2}}(\xi)$,
 $ C_1^{\frac{3}{2}}(\xi)$, $ C_2^{\frac{3}{2}}(\xi)$ are Gegenbauer polynomials,
  $\eta_3=\frac{f_{3\pi}}{f_\pi}\frac{m_u+m_d}{m_\pi^2}$ and  $\rho^2={(m_u+m_d)^2\over m_\pi^2}$
 \cite{LCSR,PSLC,Belyaev94}.

\section*{Acknowledgments}
This  work is supported by National Natural Science Foundation,
Grant Number 10405009, 10775051, and Program for New Century
Excellent Talents in University, Grant Number NCET-07-0282. The
author would like to thank Prof.T.Huang and Prof.V.M.Braun for
valuable discussions and comments.

\end{document}